\documentclass{article}
\usepackage{spconf,amsmath,graphicx}

\usepackage{amssymb}
\usepackage{graphicx}
\usepackage{url}

\usepackage{float}
\usepackage{listings}
\graphicspath{{./Figs/}}
\usepackage{graphicx}
\usepackage{graphics}
\usepackage{caption}
\usepackage{subcaption}

\DeclareGraphicsExtensions{.jpg,.eps,.png}
\usepackage{amsmath}
\usepackage{amsfonts}
\usepackage{enumerate}
\usepackage{textcomp}
\usepackage{color}
\usepackage{algorithmic}
\usepackage{algorithm}
\usepackage{amsthm}
\usepackage{bm}
\usepackage{dsfont}
\usepackage{pdfpages}
\usepackage{overpic}

\DeclareRobustCommand\onedot{\futurelet\@let@token\@onedot}

 {\everymath{\displaystyle\everymath{}}\array}%
 {\endarray}
\everymath{\displaystyle\everymath{}}

\title{Customized Facial Constant Positive Air Pressure (CPAP) Masks}
\name{{Matan Sela}, {Nadav Toledo}, {Yaron Honen},
 and {Ron Kimmel}\thanks{}}
\address{Technion - Israel Institute of Technology, 
          Haifa 32000, Israel}

\begin{document}
%
\maketitle
\begin{abstract}
Sleep apnea is a syndrome that is characterized by 
 sudden breathing halts while sleeping.
One of the common treatments involves wearing a mask that delivers continuous
 air flow into the nostrils so as to maintain a steady air pressure.
These masks are designed for an average facial model and are often difficult 
 to adjust due to poor fit to the actual patient.
The incompatibility is characterized by gaps between the mask and the face, 
 which deteriorates the impermeability of the mask and leads to air leakage.
We suggest a fully automatic approach for designing a personalized nasal
 mask interface using a facial depth scan. 
The interfaces generated by the proposed method accurately fit the geometry 
 of the scanned face, and are easy to manufacture.
The proposed method utilizes cheap commodity depth sensors and 3D printing technologies 
 to efficiently design and manufacture customized masks for patients suffering from sleep apnea. 
\end{abstract}
\begin{keywords}
Sleep Apnea, CPAP Masks, Facial Modeling, 3D Printing, Automatic Design.
\end{keywords}
\section{Introduction}
\label{sec:intro}

Sleep apnea is a disorder characterized by chronic pauses 
	in breathing.
Breathing is usually interrupted by a physical 
 block of airflow caused by 
    the soft palate, that often also leads to snoring. 
It can cause serious problems including high blood pressure, 
	mental deterioration, heart failure, sudden death, and daytime sleepiness. 
Surgical intervention, in which anatomical obstructions are removed, 
  is considered in extreme cases.
A more common treatment is creating an environment of
 continuous positive airway pressure (CPAP) to the sleeping patient.
It requires the subject to wear a mask which is connected to
 a machine that supplies the positive airflow.
 
The currently available masks fit an average face in a given 
 population or age group. 
The diversity of the geometries of human faces, especially, for 
 children with problems in birthing, poses a huge limitation to the 
 current common practice in this domain.
A mask which does not fit well, usually causes 
 the air to leak, which reduces the efficiency of the device as 
 continuous positive air pressure can not be guaranteed in this case.

In an attempt to solve this problem, we propose a fully automatic method
 for designing and manufacturing masks that best fit a given face 
 of a specific patient.
The procedure works as follows.
At a pre-processing step, we manually mark the mask contact region on
 a smooth generic facial model. 
This step is performed once for all subjects.
Next, we scan the face of the patient using a commodity range sensor 
 for generating a geometric profile and capture a color image.
Our system then automatically detects key facial feature points on the image, 
 such as the tip of the nose, the corner of the eyes, etc.
Based on these landmark points, we find a dense point-wise correspondence
 between a template model and the acquired one via a non-rigid alignment procedure.
The registration enables finding a corresponding contact region that matches
 the geometry of the scanned face.
Finally, we use the warped interface region as a set of positional constraints 
 for automatically designing an individual mask. 
The resulting mask can then be produced by printing a mold of the
  mask, and injecting medical silicone into its cavity, or directly 3D 
  printing a silicon interface.

The masks designed by our method optimally match the scanned face and 
 do not require manual intervention during design and manufacturing of the masks.
To evaluate the quality of the new design, we estimate the force variations along the contact
 region between the mask and the face. 
Unlike standard interfaces, the pressure using the proposed new design  
  is uniformly distributed along the contact contour between 
  the mask and the face.
The result is convenient, compact, and efficient devices 
 at a relatively insignificant additional cost. 

\begin{figure*}[ht]
	\begin{center}
	\begin{overpic}[width=1.0\textwidth]{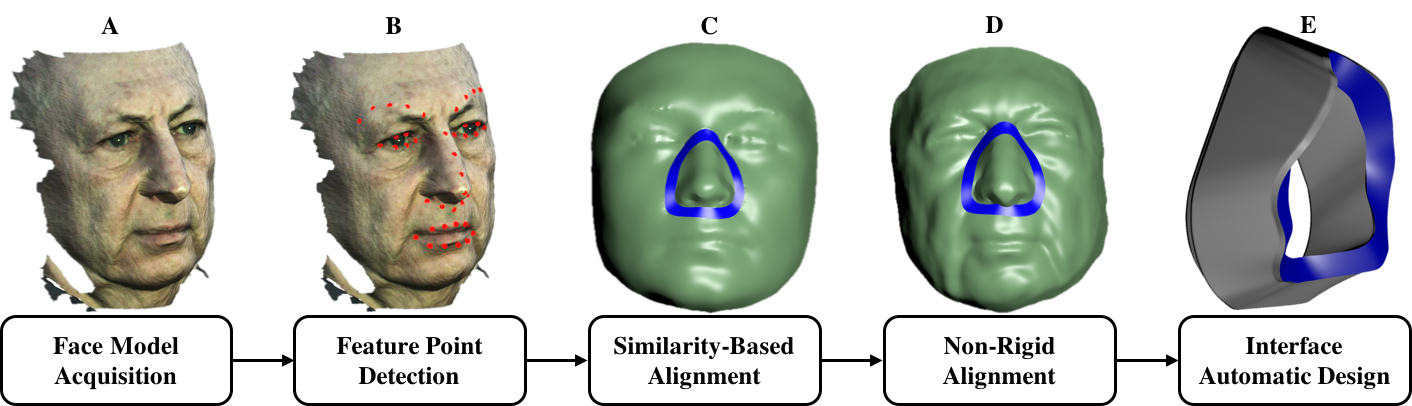}
	\end{overpic}
	\end{center}
	\caption{\small 
	The pipeline of the proposed method: 
    First, a three dimensional model of the face 
    using a depth sensor is acquired (A).
	Key characteristic points are 
	then detected on the face
	such as the nose center,
	eye lashes, etc. (B).
	Based on these feature points,
	a generic template face is aligned 
	with the scan (C).
	This registration enables
	automatic detection of the mask contact region 
    upon the scanned face (D). 
	We then automatically design a mask that 
    matches the facial features (E).}
	\label{fig:blockdiagram}
\end{figure*}
\section{Related Efforts}
\label{sec:related_effort}
Several recent papers introduce attempts to simplify and partially automate
 the customization of personalized masks.
Cheng et al. \cite{cheng2013application} proposed to reconstruct a three dimensional  
 facial model of the patient by immersing the face into Hygrogum material. 
After solidifying, the  Hygrogum turns into a sub-millimeter accurate mold.
Morrison et al. \cite{morrisonpersonalized} and 
 Cheng et al. \cite{cheng2015clinical} proposed to use a depth 
 scanner for facial modeling.
Amirav et al. \cite{amirav2014design}
 scanned hundreds of newborns, clustered them into three 
  typical sets, and used a representative from each set 
  to design three types of inhalation masks for babies.
After digitizing the geometry of the face, almost all
 existing methods project the profile
 of a given mask onto the reconstructed facial surface 
 for customizing the mask interface. 
 Unlike existing methods, the proposed solution 
  allows determining the contact region based on 
  warping a pre-design model.
The shape of the mask depends on the patient's face,
 and it is designed to smoothly
 match the facial contours of the patient.
 

\section{Implementation Considerations}
\label{sec:implementation}

\noindent{\bf Overview:}
	A sketch of the proposed procedure is shown in 
	Figure \ref{fig:blockdiagram}. 
    Next, we describe  each part of the system.\\

\noindent{\bf Face Model Acquisition:}
Range scanners measure the distance of each
 visible point in the scene from the sensor.
There are various methods for capturing range images.
Here, we used a structured light sensor which is
 comprised of a calibrated projector and a camera. 
The projector marks each point in the scene with
 a unique signature. 
It allows the camera, which is located at a known 
 relative position, to identify each visible point.
Then, a simple triangulation is performed for 
 estimating the distance of each point from
 the camera.
	
We used a single depth scan acquired by a color
 structured light sensor, as the accuracy of 
 this system is sufficient for our purpose 
  as shown in Figure 1A.
Alternatively, it would be possible to acquire 
 multiple frames using off-the-shelf depth sensors, 
  such as Intel's Realsense, and to fuse them 
  into a single three dimensional facial model 
  via techniques such as those proposed in
  \cite{kinfu12} or \cite{Curless1996}. 
These methods reduce the noise in each single scan 
 by fusing many scans together into a refined model.\\
    
\noindent{\bf Feature Points Detection:}
For initializing the alignment procedure, our system
 finds a small set of predefined characteristic points 
 on the scanned face. 		
These points can be detected on a color image using Active 
	Shape Model algorithms such as \cite{cootes1995active}.
In practice, we estimate the position of approximately 
 sixty landmarks on the face, see Figure 1B.
To ensure robustness to outliers, we remove feature points 
 detected at pixels whose depth value was not evaluated.\\

\noindent{\bf Similarity Initial Alignment:}
Based on the detected landmark points, we estimate the similarity
 transformation applied to the template face, see Figure 1C, 
  which minimizes the distances between corresponding feature 
  points on the template and the target face. 
Notice that throughout the alignment process, the scanned 
 facial model remains static and only the template face is
 transformed.  

Denote the set of feature points detected on the scan as 
 $\{ r^{scan}_1,...,r^{scan}_k \}$, and the set of 
 corresponding feature points on the template face, which 
 we selected in advance, as $\{r^{temp}_1,...,r^{temp}_k\}$.
For initializing the scaling factor $\alpha$, we minimize 
 the term
\begin{align*}
\min_{\alpha} \sum_{i,j}^k \|\alpha  
   \cdot d(r^{temp}_i,r^{temp}_j) - d(r^{scan}_i,r^{scan}_j)\|^2_2.
\end{align*}
Here, $d(\cdot,\cdot)$ represents the Euclidean distance
 between a couple of points.

Next, we calculate the rotation matrix $R$, the translation
 vector $t$ and update the scaling factor $\alpha$ iteratively,
 by minimizing 
\begin{align*}
\min_{R \in so(3),t \in \mathds{R}^3 ,\alpha \in \mathds{R}_+} \sum_{i=1}^{k}\|(\alpha R r^{temp}_i + t) -r^{scan}_i\|^2 
\end{align*}
The optimization is done by an alternating minimization of the
 three variables.
The process converges after few iterations with an accurate
 similarity transformation that we use for initializing
 the non-rigid alignment algorithm described next.\\

\noindent{\bf Non-Rigid Alignment:}
At a preprocessing step, we mark the region of contact of the
 mask on the generic face, see Figure 1C.
In order to find a corresponding region on the scanned subject, 
 we elastically deform the template face until it is aligned 
 with the scanned one, see Figure 1D.
The elastic energy we minimize penalizes local stretching and
 bending. 
Thus, the resulting registration is an almost isometric map 
 between the faces.

The procedure is performed in an iterative closest point fashion.
Each iteration involves four steps. 
First, we find for each vertex of the template face an 
 approximate nearest neighbor on the reconstructed one. 
Next, we eliminate pairs from the obtained correspondence list 
 which are too far from one another or have normals with 
 different orientations. 
Based on the remaining pairs in the list, we then estimate 
 the displacement of each point of the template face, by 
 minimizing an energy term we describe next.
Finally, we check the norm of the motion between the 
 last consecutive iterations and, if necessary, readjust the 
 weights in the objective function for the next iteration.

In the nearest neighbor matching step, for each point of
 the template face $v_i^{temp}$, a close point on the 
 scanned surface $c_i^{scan}$ is found. 
First, we  construct a KD-tree for the point cloud of the 
 scanned surface. 
The KD-tree  is constructed by recursively splitting a 
 volumetric cell along the $x$, the $y$, or the $z$ axes, 
 into two subcells where each contains half of the points. 
The recursion stops when it reaches a leaf cell which contains 
 a single point. 
For finding the nearest neighbor of a query point, we travel 
 along the KD-tree until we reach a leaf corresponding to
 the approximate nearest neighbor.
In cases where the depth image of the subject is available, we
 find the nearest neighbor by projecting each template point 
 onto the scan image grid and associate it with the point 
 that belongs to the nearest pixel. 
This reduces the complexity from $O(nlogm)$ for the KD-tree 
 to $O(n)$, where, $n$ denotes the number of queried points 
 and $m$ is the number of points on the scanned surface.

The obtained list of point correspondences 
 $\{(v_i^{temp},c_i^{scan} )\}_{i=1}^p$ may include outliers. 
This can introduce undesired artifacts in the deformation 
 procedure that follows, as the scanned surface may have
 holes and noise. 
Thus, we  remove matching pairs which are more than five 
 millimeters apart, and pairs whose normal directions differ
 at more than twenty five degrees. 

Based on the remaining matching pairs
 $\{(v_i^{temp},c_i^{scan} )\}_{i=1}^{p'}$, we elastically deform the 
 template face. 
The deformation is modeled by a displacement field 
 $\underbar{d} = (d_1^{temp},...,d_{n_{temp}}^{temp})$. 
Namely, we optimize over the change in position of each point 
 on  the template face, such that 
  $v_i^{temp} \leftarrow v_i^{temp} + d_i^{temp}$. 
The displacement is found by minimizing the following
\begin{align*}
 E(\underbar{d}) = & \alpha_{p2point} \cdot E_{p2point}(\underbar{d})  \cr
&  + \alpha_{p2plane} \cdot E_{p2plane}(\underbar{d})  \cr
& +\alpha_{memb} \cdot E_{memb}(\underbar{d})  \cr
& + \alpha_{ref} \cdot E_{ref}(\underbar{d}),
\end{align*}
where $\alpha _{(\cdot)}$  are positive scalar weights and the  
 energy terms are given by 
\begin{itemize}
\item Point-to-point energy - The sum of squared Euclidean 
 distances between corresponding points in the list
\begin{align*}
E_{p2point}\left(\underbar{d}\right)= \sum_{i=1}^{p'} \| v^{temp}_i + d_i^{temp} - c^{scan}_i \|^2.
\end{align*}
\item Point-to-plane energy - the sum of squared Euclidean
 distances between a point on the template and the tangent
 plane of the corresponding point on the scanned surface
\begin{align*}
E_{p2plane}  \left(\underbar{d}\right) &= \cr  & \sum_{i=1}^{p'} \left| n^{scan}_i \cdot \left(v^{temp}_i + d_i^{temp} - c^{scan}_i\right)^T \right|^2,
\end{align*}
 where $n^{scan}_i$ is the unit normal at the vertex $c^{scan}_i$.
\item Biharmonic energy - this regularization term enforces 
 the smoothness of the displacement field as functions on
 the template face.
\begin{align*}
E_{memb}(\underbar{d}) & = \cr & \sum_{i=1}^{n_{temp}} \sum_{j \in \mathcal{N}\left(v^{temp}_i\right)} \|w_{i,j} \left(d_i^{temp} - d_j^{temp} \right) \|^2,
\end{align*}
 where $w_{i,j}$ are the cotangent weights, see e.g. \cite{meyer03} 
 for more details, and $\mathcal{N}\left(v^{temp}_i \right)$ 
 are the set of neighboring vertices of $v^{temp}_i$. 
For more details about this enrergy, we refer to 
 \cite{Welch1992} and \cite{botsch2008linear}.
\item Constraint energy - this term measures the sum of 
 squared Euclidean distances between the detected 
 corresponding feature points.
\begin{align*}
E_{ref}(\underbar{d}) = \sum_{i=1}^k \|r^{temp}_i + 
                    d_i^{temp} - r^{scan}_i\|^2.
\end{align*}
\end{itemize}

Each of the above energy terms is weighted differently in 
 the objective function. 
To make the registration robust to local minima, we perform 
 the alignment gradually in a coarse-to-fine fashion. 
At the first iteration we set $\alpha_{p2point} = 0.1$, 
 $\alpha_{p2plane} = 1$, $\alpha_{memb}  = 100$ and 
 $\alpha_{ref} = 10$. 
At the end of each iteration, we measure the norm of the
 displacement field relative to the previous iteration. 
If the value is below $10^{-2}$, we decrease $\alpha_{memb} $ 
 and $\alpha_{ref}$ by half.
The algorithm converges after $10-20$ iterations with an 
 accurate and smooth alignment. 
\\

\noindent
{\bf Automatic Interface Design:}
To customize a mask interface which fits the geometry of the 
 patient with smooth contact boundaries, we model the interface 
 with non-uniform rational basis spline (NURBS). 
This tool is frequently used in computer graphics for modeling 
 shapes as it provides an intuitive design framework. 
The surfaces modeled using NURBS can be modified by editing
 the position of control points based on which the surface
 is smoothly interpolated. 
For more details about this technique, we refer to 
 \cite{piegl2012nurbs}.

For fitting the mask to the warped contact region, we first 
 place the generic mask interface next to the warped model, 
 and then modify the position of control points along the 
 side facing the model, see Figure 1E). 
We use $256$ control points which we translate to the position
 of the corresponding points on the warped contact region.
The correspondence between the control points and the contact
 region is manually determined a priori.

\section{Results}
The generic template face we used contains approximately $36,000$
 points. 
The initial mask interface was manually designed using CAD tools
 and has roughly $20,000$ points. 
For the automatic interface design we used a Python script coded
 in Blender which uses its NURBS tool.
The resulting mask interface is shown in Figure 1E.

\begin{figure}[ht]
	\begin{center}
	\begin{overpic}[width=0.7\columnwidth]{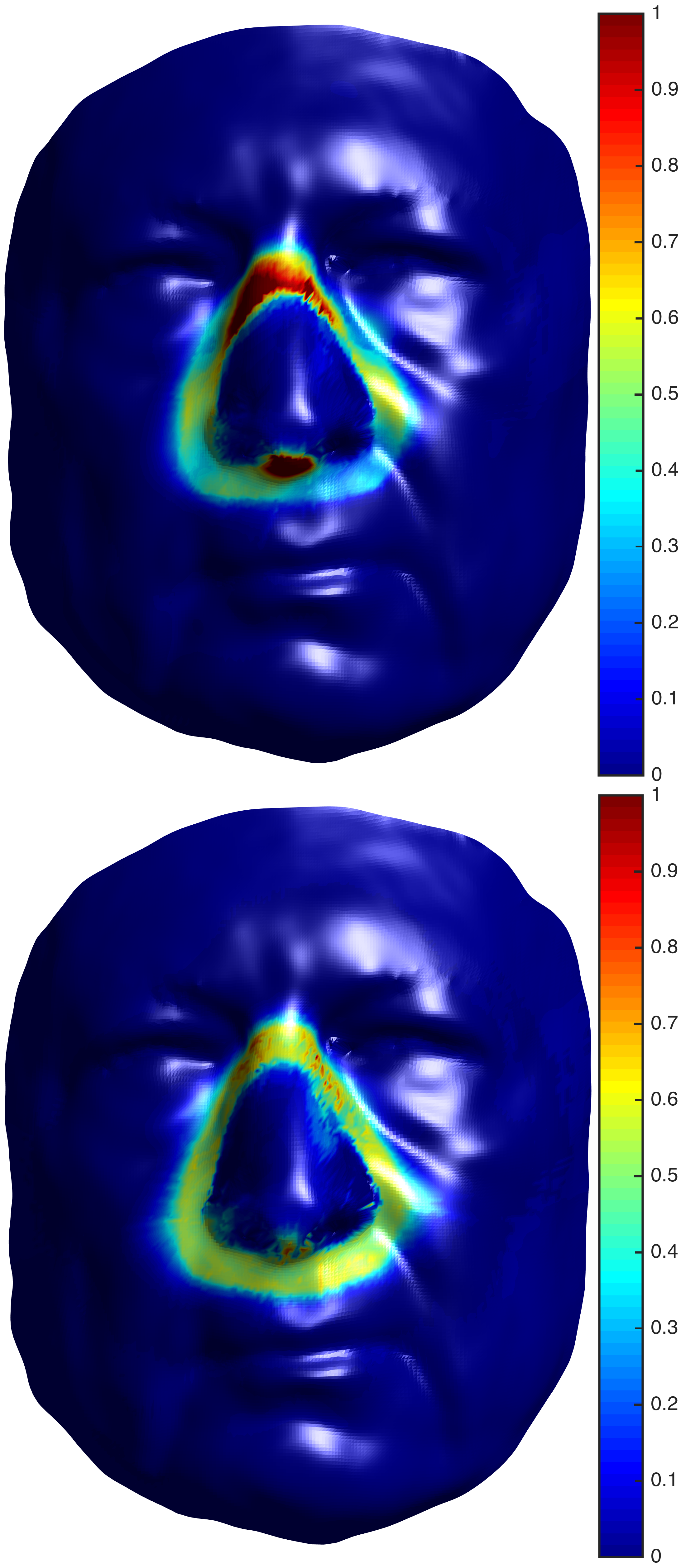}\end{overpic}
	\end{center}
	\caption{\small Pressure simulation of commodity 
     mask pressed onto a facial surface (top) and a  
      personalized mask designed by the proposed method (bottom).}
	\label{fig:force_sim}
\end{figure}
For evaluating the mask interface fitting to the scanned face, 
 we performed an experiment for estimating the force 
 distribution along the contact rim. 
The face was modeled as a soft physical body with 
 a friction of $50$.
We then simulated a collision between the mask interface and
 the face using Blender's physical simulation. 
Next, we calculated the norm of the relative motion between 
 each point on the face before and after the simulation. 
We repeated this experiment twice, once for the mask that
 was designed for an average facial model and once for 
 a mask designed for the specific patient.
The results are shown in Figure \ref{fig:force_sim}. 
The pressure along the contact region generated by the 
 personified mask is shown to spread more evenly compared to 
 the generic mask designed for an average facial model.

\section{Conclusions}
We introduced a method for an automatic CPAP mask 
 modeling given a facial depth scan. 
A generic face with predefined mask contact contour 
 is matched to the scanned model using an iterative
 elastic registration method. 
The warped region is then used for setting constraints
 for the mask interface design.
The resulting mask fits the facial features of the  
 patient making the mask impermeable, more efficient, 
 and more convenient compared to existing designs. \\
 

\noindent
{\bf Acknowledgements:}
This work has been supported by advanced  FP7-ERC grant
 No. 267414.
The authors thank Mr. Alex Porotsky for participating
 in the experiments.
\clearpage
{\small
\bibliographystyle{IEEEbib}
\bibliography{strings,refs}
}
\end{document}